\documentclass[review]{elsarticle}

\usepackage{amsmath,amssymb}

\usepackage{graphics} 
\usepackage{epsfig} 
\usepackage{times} 
\usepackage{amssymb}  
\usepackage{adjustbox}
\usepackage{soul, color} 
\usepackage[english]{babel}
\usepackage{graphicx}

\usepackage{subcaption}
\usepackage{diagbox}
\usepackage{bbm}

\usepackage{lmodern,babel,adjustbox,booktabs,multirow}

\usepackage{upquote}
\usepackage{mathrsfs}
\usepackage{lmodern}
\usepackage{bm}
\usepackage{algorithm}
\usepackage{algorithmic}
\usepackage{booktabs,caption}
\usepackage[flushleft]{threeparttable}
\usepackage{multirow}
\usepackage{mathrsfs}
\usepackage{adjustbox}

\usepackage{xpatch}
\usepackage{nomencl}

\makeatletter
\def\ps@pprintTitle{%
 \let\@oddhead\@empty
 \let\@evenhead\@empty
 \def\@oddfoot{\centerline{\thepage}}%
 \let\@evenfoot\@oddfoot}

\xpatchcmd{\thenomenclature}{%
  \section*{\nomname}
}{
}{\typeout{Success}}{\typeout{Failure}}
 
\begin{document}

\begin{frontmatter}

\title{Data-driven sparse polynomial chaos expansion for models with dependent inputs}

\author{Zhanlin Liu\fnref{myfootnote}}

\author{Youngjun Choe\fnref{myfootnote2}}


\begin{abstract}
Polynomial chaos expansions (PCEs) have been used in many real-world engineering applications to quantify how the uncertainty of an output is propagated from inputs. PCEs for models with independent inputs have been extensively explored in the literature. Recently, different approaches have been proposed for models with dependent inputs to expand the use of PCEs to more real-world applications. Typical approaches include building PCEs based on the Gram-Schmidt algorithm or transforming the dependent inputs into independent inputs. However, the two approaches have their limitations regarding computational efficiency and additional assumptions about the input distributions, respectively. In this paper, we propose a data-driven approach to build sparse PCEs for models with dependent inputs. The proposed algorithm recursively constructs orthonormal polynomials using a set of monomials based on their correlations with the output. The proposed algorithm on building sparse PCEs not only reduces the number of minimally required observations but also improves the numerical stability and computational efficiency. Four numerical examples are implemented to validate the proposed algorithm.   
\end{abstract}

\begin{keyword}
uncertainty quantification \sep polynomial chaos expansion \sep sparse polynomial chaos expansion \sep Gram-Schmidt orthogonalization 
\end{keyword}

\end{frontmatter}


\section{Introduction}
Uncertainty quantification plays a critical role in many domains of real-world engineering applications as it characterizes the uncertainties of the system outputs in those applications. Surrogate models often serve as mathematical models to describe how the uncertainty of a system output is propagated from the inputs. Among the surrogate models, polynomial chaos expansions (PCEs) have been widely used to 
conduct the uncertainty quantification on the outputs in many industrial applications including thermodynamics \cite{avdonin2018uncertainty}, electromagnetism \cite{chen2015sensitivity}, chemical engineering \cite{xie2019efficient}, aerodynamics
\cite{palar2018global}, , hydrogeology \cite{deman2016using}, structural safety analysis \cite{xu2019structural, schobi2019global}, power systems \cite{prempraneerach2010uncertainty}, and manufacturing \cite{liu2020identifying, hawchar2017principal}.

To accurately conduct uncertainty quantification for models with different types of inputs, a variety of PCEs have been developed in the literature. For models with independent inputs, the Wiener chaos expansion, which is known as the first PCE in the literature, uses Hermite polynomials to construct PCE models for Gaussian-distributed inputs \cite{Wiener:1938}. Later PCEs, including the generalized PCE (gPCE) \cite{Xiu:2002}, the multi-element gPCE (ME-gPCE) \cite{Xiaoliang:2006}, the moment-based arbitrary PCE (aPCE) \cite{Oladyshkin:2012}, the support vector regression based PCE \cite{cheng2018adaptive}, and the Gram-Schmidt based PCE (GS-PCE) \cite{Witteveen:2007}, are developed for independent inputs following non-Gaussian distributions.

Even though the GS-PCE can also be used to construct PCEs for models with dependent inputs, the procedure of using Gram-Schmidt algorithm is computationally demanding as the number of the inputs increases or the expansion order increases \cite{torre2019data}. Therefore, \cite{torre2019data} provides an alternative method to construct PCE for models with dependent inputs by transforming the dependent inputs into independent inputs using the Rosenblatt transformation. However, this approach might not be applicable to many engineering applications due to the fact that the Rosenblatt transformation requires the knowledge about the conditional probability density functions about the inputs, which is often not the case in practice. Thus, how to efficiently construct a PCE for models with dependent inputs without using distribution information about the inputs still requires more investigations.

Consequently, the main contribution of this paper is to propose a sparse PCE algorithm which can efficiently construct sparse PCEs for models with both independent or dependent inputs regardless of the input distribution. To the best of our knowledge, the proposed method is also the first data-driven method that constructs a sparse PCE for models with dependent inputs without requiring a large number of observations. We validate the proposed method empirically using four simulation examples by estimating the variance of the output and the Kullback-Leibler (KL) divergence from the estimated output distribution to the true distribution. The simulation examples show the advantage of the proposed method in terms of computational speed and numerical stability of constructing orthogonal polynomials \cite{torre2019data}. In addition, the proposed method is more accurate on estimating the lower-order moments and distribution of the output comparing with the state-of-the-art method on constructing sparse PCE models regardless of the input distribution and their dependency structure.

The remainder of this paper is organized as follows. Sec.~\ref{sec:back} briefly reviews the technical background on the PCE and the state-of-the-art methods on constructing sparse PCEs. Sec.~\ref{sec:method} proposes the algorithm of the proposed method and discusses the advantages of using the proposed method. In Sec.~\ref{sec:appli}, the proposed algorithm is empirically evaluated using four simulation examples. 
Sec.~\ref{sec:conclusion} concludes the paper with a discussion on future research directions. 

\section{Background}
\label{sec:back}
In this section, we will first introduce PCE models and how to estimate the lower-order moments using the PCE model coefficients. Then we will review how to construct GS-PCE since it is regarded as the pioneering work of data-driven PCE model regardless of the distribution and dependency of the inputs. In the end, we will review the state-of-the-art routine for constructing sparse PCEs for dependent inputs, which applies the least angle regression method on orthonormal polynomials constructed using the modified Gram-Schmidt algorithm.

\subsection{PCE model}
\label{sec:PCE}
PCE models the relationship between the $n$ random inputs in $\boldsymbol X$ and the output $Y$ using a finite number of orthonormal polynomials as follows:
\begin{equation}
\label{eq:exp}
Y=f(\boldsymbol X) \approx \sum_{i=0}^{P}\theta_{i}\psi_{i}(\boldsymbol X),
\end{equation}
where $\theta_{i}$, $i$ = 0,1,2,\ldots,$P$, are called PCE coefficients and $\psi_{i}$, $i$ = 1,2,\ldots,$P$ are orthonormal polynomials. The orthonormal polynomials can be constructed based on different PCE models. We will particularly introduce how to construct orthonormal polynomials using the modified Gram-Schmidt algorithm in Section~\ref{sec:GS-PCE}.  
\begin{equation}
\label{eq:exp2}
P +1 = \binom{n+p}{n}
\end{equation}
is the number of polynomial terms, where
$p$ is the highest polynomial degree in the PCE model. As $p$ increases to infinity, the error of estimating the output using the PCE model converges to $0$ \cite{Cameron:1947}.

In this paper, the PCE coefficients are solved by solving an overdetermined linear system equations in the least-squares sense using a regression as follows \cite{pettersson2015polynomial}:
\begin{equation}
\begin{aligned}
\label{eq:over_lin}
 & \underset{\boldsymbol{\theta}\in\mathbb{R}^{P+1}}{\mathrm{argmin}} \sum_{j=1}^{m}\left(Y_{j} - 
 \sum_{i=0}^{P}\theta_{i}\psi_{i}\left(\boldsymbol{X}_{j}\right)\right)^{2}, \\ 
\end{aligned}
\end{equation}
where $\boldsymbol{\theta}$ denotes $\left(\theta_{0}, \theta_{1},\ldots, \theta_{P}\right)$. $Y_{j}$ and $\boldsymbol{X}_{j}$ represent the output and the input vector of the $j^{th}$ observation, $j=1,\ldots,m$, respectively. 


Thanks to the orthogonality of the orthonormal polynomials, we can approximate the lower-order moments of output $Y$ using the PCE coefficients as follows: 
\begin{equation}
\begin{aligned}
\label{eq:mom}
&\mathbb{E}(Y) \approx \theta_{0},\\
&Var(Y) \approx \sum_{i=1}^{P}\theta_{i}^2.
\end{aligned}
\end{equation}
The accuracy of estimating the lower-order moments improves as $P$ in Eq.~\eqref{eq:mom} increases.


\subsection{GS-PCE}
\label{sec:GS-PCE}
The GS-PCE constructs orthonormal polynomials based on $P$ initial polynomials $\left(e_{i}\right)_{i \in\{1,2,\ldots, P\}}$, where $e_{i}, i=1,2,\ldots, P,$ are assumed to be linearly independent. Then the orthonromal polynomials $(\psi_{i}(\boldsymbol{X}))_{i \in\{1,2,\ldots, P\}}$ are obtained using the modified Gram-Schmidt algorithm described as follows \cite{liu2018data}:
\begin{algorithm}[H]
\caption{Modified Gram-Schmidt Algorithm}
\begin{algorithmic}[1] \label{Gram-Schmidt} 
\REQUIRE $P$ linearly independent initial polynomials $\left(e_{i}\right)_{i \in\{1,2,\ldots, P\}}$.
\\
\ENSURE $P$ orthonormal polynomials $(\psi_{i}(\boldsymbol{X}))_{i \in\{1,2,\ldots, P\}}$.
     \FOR{$i = 1, 2, \ldots, P $}
	\STATE $\phi_{i}(\boldsymbol X) \leftarrow e_{i}(\boldsymbol 	X)$
	\FOR{$k = 1, 2, \ldots, i-1 $}
\STATE{$\phi_{i}(\boldsymbol X) \leftarrow  \phi_{i}(\boldsymbol X) - \langle \phi_{i}(\boldsymbol X), \psi_{k}(\boldsymbol X) \rangle \psi_{k}(\boldsymbol X) $}
      	\ENDFOR
 	\STATE $\psi_{i}(\boldsymbol X)  \leftarrow\frac{\phi_{i}(\boldsymbol X)}{||\phi_{i}(\boldsymbol X)||_{2} }$
	\ENDFOR
\end{algorithmic}
\end{algorithm}
\noindent The inner-product in the algorithm is defined with respect to the empirical measure in this paper. 

Even though GS-PCE provides feasibility on constructing orthonormal polynomials for dependent inputs following arbitrary distributions, it is computationally demanding as the number of input or polynomial order increases \cite{torre2019data}. In addition, the GS-PCE might be inaccurate for models with highly correlated inputs since the constructed orthonormal polynomials might lose their orthogonality due to the rounding error \cite{giraud2005rounding}. 

\subsection{Sparse PCE}\label{subsec:sparsePCE}
A variety of sparse PCEs have been explored in the literature. \cite{blatman2008sparse} proposes a greedy forward-backward selection algorithm, which is regarded as a pioneering work for constructing sparse PCEs in the literature. Based on this work, many other techniques have been introduced to construct sparse PCEs using the least angle regression (LAR) and the diffeomorphic modulation under observable response preserving homotopy (D-MORPH) regression \cite{cheng2018sparse} as well as solving a sparse regression with a regularization term \cite{jakeman2015enhancing, guo2018gradient, zhou2020active}. These techniques can also be applied to construct sparse PCEs with $m$ orthonormal polynomials using the algorithm summarized in Algorithm \ref{alg:lar}. Even though there also exist other methods for constructing sparse PCEs without using the Gram-Schmidt algorithm \cite{blatman2010efficient, pan2017sliced, zhou2020active, lim2021distribution}, they require assumptions on the dependency or distributions of the inputs. 

\begin{algorithm}[h]
\caption{Sparse PCE algorithm for models with dependent inputs}
\begin{algorithmic}[1]\label{alg:lar} 
\REQUIRE At least $P+1$ random observations of output $Y$ and inputs $\boldsymbol{X}$.
\\
\ENSURE A sparse PCE representation of $Y$ with respect to $\{\psi_{i^{\prime}}(\boldsymbol{X})\}_{i^{\prime}=1}^{m}$. 
\STATE{Construct $P$ initial linearly independent polynomials $(e_{i})_{i \in \{1,2,\ldots,P\}}$. }
\STATE{Construct orthonormal polynomial basis $\{\psi_{i}(\boldsymbol{X})\}_{i=1}^{P}$ using the modified Gram-Schmidt polynomials described in Algorithm \ref{Gram-Schmidt}.}
\STATE{Construct a sparse PCE model of $Y$ by selecting $\{\psi_{i^{\prime}}(\boldsymbol{X})\}_{i^{\prime}=1}^{m} \subseteq \{\psi_{i}(\boldsymbol{X})\}_{i=1}^{P}$ based on a sparse regression.  }
\end{algorithmic}
\end{algorithm}
As the procedure of the modified Gram-Schmidt algorithm is embedded in Algorithm \ref{alg:lar}, it inherits the computational inefficiency and inaccuracy from Algorithm \ref{Gram-Schmidt}. To address these drawbacks, we propose an algorithm in the following section that improves the efficiency and accuracy of constructing a sparse PCE for a model with dependent inputs. Step 3 in Algorithm \ref{alg:lar} can use any sparse regression method. In this paper, as the benchmark method for empirical validation in Section~\ref{sec:appli}, we use the LAR-based method to build sparse PCEs.

\section{Methodology}
\label{sec:method}

As we described in Section~\ref{sec:GS-PCE}, constructing orthonormal polynomials becomes more computationally demanding as the number of inputs or polynomial order increases. Therefore, we propose a new algorithm which builds a sparse PCE regardless of the distributions and dependency structure of the inputs. The proposed algorithm not only removes the need for a large number of random observations but also improves the estimation accuracy and computational speed.

Unlike the benchmark method described in Algorithm \ref{alg:lar}, which constructs $P$ orthonoraml polynomials based on $P$ initial linearly independent polynomials before applying an operator to construct a sparse PCE, the proposed algorithm only constructs a limited number of orthonormal polynomials that significantly explain the output $Y$ from $P$ initial polynomials. 

The proposed sparse PCE algorithm is a recursive algorithm which requires a pre-defined threshold value $\epsilon \in (0,1)$ and a set of initial polynomials $\{e_{i}^{(0)}\}_{i \in \{1,2,\ldots,P^{(0)}\}}$ at its initial step, where $l=0$ represents the iteration counter. As it is defined in Eq.~\eqref{eq:exp2}, $P^{(0)}$, which is the number of the initial polynomials, depends on the number of inputs $n$ and the polynomial order $p$. In this paper, the $P^{(0)}$ initial polynomials are constructed by the tensor product of the univariate polynomial of each input $X_{i}$ in $\boldsymbol{X}, i=1,2,\ldots, n$ as follows:
\begin{equation}
\label{eq:poly}
\{e_{i}^{(0)}\}_{i \in \{1,2,\ldots,P^{(0)}\}} = \left\{ \prod_{k=1}^{n}X_{k}^{j_{k}} : 
j_{k} \in \{0,1,\ldots p\}, \sum_{k=1}^{n}j_{k} \leq p\right\}. 
\end{equation}

In the $l^{th}$ iteration, we calculate $\forall i \in \{1,2,\ldots,P^{(l-1)}\}, \rho(e_{i}^{(l-1)}(\boldsymbol{X}), Y)$, where $\rho(\cdot,\cdot)$ is an operator that calculates the empirical Pearson correlation coefficient between two variables. The algorithm stops with the following condition:
\begin{equation}
    \label{eq:thre_1}
    \forall i \in \{1,2,\ldots,P^{(l-1)}\}, \rho(e_{i}^{(l-1)}(\boldsymbol{X}), Y) < \epsilon.
\end{equation}
Otherwise, we create $\{e_{i}^{(l)}\}_{i \in \{1,2,\ldots,P^{(l)}\}}$ by keeping \textit{only} $e_{i}^{(l-1)}(\boldsymbol{X})$ 
that satisfies the condition 
\begin{equation}
    \label{eq:thre_2}
  \rho((e_{i}^{(l-1)})(\boldsymbol{X}), Y) \geq \epsilon, \forall i \in \{1,2,\ldots,P^{(l-1)}\}. 
\end{equation}
This procedure prevents selecting polynomials that are linearly dependent or highly correlated with the constructed orthonormal polynomials in the previous iterations and reduces the number of constructed orthonormal polynomials compared with the modified Gram-Schmidt algorithm. In the next step, we select one polynomial ($e_{i}^{(l)}(\boldsymbol{X}) \in \{e_{i}^{(l)}\}_{i\in \{1,2,\ldots, P^{(l)}\}}$), which satisfies the equation as follows:
\begin{equation}
    \label{eq:select}
    \forall j \in \{1,2,\ldots,P^{(l)}\}, \rho(e_{i}^{(l)}(\boldsymbol{X}), Y) \geq \rho(e_{j}^{(l)}(\boldsymbol{X}), Y).
\end{equation}
After we get $e_{i}^{(l)}(\boldsymbol{X})$, we transform it into an orthonormal polynomial with respect to $\{\boldsymbol{\psi}_{i}(\boldsymbol{X})\}_{i=1}^{l-1}$ using Steps 2--6 in Algorithm \ref{Gram-Schmidt}. Unlike the modified Gram-Schmidt algorithm, which constructs orthonormal polynomials based on the initial polynomials whose ordering is defined, the proposed algorithm constructs the orthonormal polynomials based on their correlations with $Y$. To be more specific, the polynomials which have high correlations with $Y$ are selected first to construct the orthonormal polynomials. It is shown that such ordering improves the numerical stability of constructing orthonormal polynomials in Section~\ref{sec:appli}. The proposed algorithm is summarized in Algorithm \ref{algorithm_sf}.

\begin{algorithm}[ht]
\caption{Forward-selection sparse PCE (FSS-PCE) algorithm}
\begin{algorithmic}[1]\label{algorithm_sf} 
\REQUIRE Random observations of output $Y$ and inputs $\boldsymbol{X}$; Threshold value $\epsilon$; Iteration counter $l=0$. 
\\
\ENSURE A sparse PCE representation of $Y$ with respect to $\{\boldsymbol{\psi}_{i}(\boldsymbol{X})\}_{i=0}^{l}$. 
\STATE{Construct $P$ initial polynomials $\{e_{i}^{(l)}\}_{i \in \{1,2,\ldots,P^{(l)}\}}$ using Eq.~\eqref{eq:poly}. }
\STATE{Increase $l$ by 1 and select $e_{l}(\boldsymbol{X}) \in \{e_{i}^{(l-1)}\}_{i \in \{1,2,\ldots,P^{(l-1)}\}}$ based on Eq.~\eqref{eq:select}.}
\STATE{Update $\{e_{i}^{(l)}\}_{i \in \{1,2,\ldots,P^{(l+1)}\}}$ based on Eq.~\eqref{eq:thre_2}. }

\STATE{Transform $e_{l}(\boldsymbol{X})$ into $\psi_{l}(\boldsymbol{X})$ using Steps 3--6 in Algorithm \ref{Gram-Schmidt}, where $e_{l}(\boldsymbol{X})$ and $\{\boldsymbol{\psi}_{i}(\boldsymbol{X})\}_{i=0}^{l-1}$ replace $\phi_{i}(\boldsymbol{X})$  and $\{\psi_{k}(\boldsymbol{X})\}_{k=1}^{i-1}$, respectively.}
\STATE{Check Eq.~\eqref{eq:thre_1} and if satisfied, go to step 6. Otherwise, go to Step 2. }
\STATE{Model the output $Y$ using a sparse PCE model with respect to $\{\boldsymbol{\psi}_{i}(\boldsymbol{X})\}_{i=1}^{l}$ using Eq.~\eqref{eq:exp}}.
\end{algorithmic}
\end{algorithm}

The pre-specified $\epsilon$ decides the goodness of fitting and the sparseness of the PCE model. It is chosen by conducting an $K$-fold cross-validation. To conduct the $K$-fold cross-validation, we first randomly split all random observations into $K$ subsets where each subset contains the same number of observations. Then we treat each subset as a testing set and the rest of the subsets as a training set. After that, we construct a sparse PCE using Algorithm \ref{algorithm_sf} and estimate PCE coefficients of a sparse PCE based on the training set and predict $Y$ on the testing set using its inputs. By following this procedure for $K$ times, where each time we use a different subset as the testing set, we choose $\epsilon$ based on the optimization as follows: 
\begin{equation}
\begin{aligned}
    \label{eq:cross-vali}
     & \underset{\epsilon\in(0,1)}{\mathrm{argmin}}\sum_{j=1}^{n}\left(\boldsymbol{Y}_{j} - 
 \sum_{i=0}^{P^{(j)}}\theta_{i,\epsilon}^{(j)}\psi_{i,\epsilon}^{(j)}\left(\boldsymbol{X}_{j}\right)\right)^{2}, \\ 
\end{aligned}
\end{equation}
where $\boldsymbol{Y}_{j}$ represents the outputs in the $j^{th}$ fold. $\{\psi_{i}^{(j)}\left(\boldsymbol{X}_{j}\right)\}_{i=0}^{P^{(j)}}$ and $\boldsymbol{\theta}^{(j)}$ are the orthonormal polynomials and the PCE coefficents estimated based on the rest of the folds, respectively.

\section{Empirical validation}
\label{sec:appli}
In this section, we present four numerical examples to empirically validate the proposed method. The first and second examples consider the inputs that are independent and dependent, respectively, in synthetic settings. The third and forth examples consider modeling the output using sparse PCE for dependent inputs in real-world problems.  

In this paper, we use the relative error (RE) to compare the accuracy of estimating the standard deviation of $Y$ for both the benchmark method and the proposed method. The relative error is defined as follows:
\begin{equation}
    \label{eq:reerro}
    \epsilon_{re} = \frac{|\sigma_{Y} - \hat{\sigma}_{Y}|}{\sigma_{Y}},
\end{equation}
where $\sigma_{Y}$ is the theoretical standard deviation of $Y$ or an estimate using the Monte Carlo method based on a large number of random observations. $\hat{\sigma}_{Y}$ is the estimated standard deviation of $Y$ using either the benchmark method or the proposed method using Eq.~\eqref{eq:mom}. In each simulation example, a smaller $\epsilon_{re}$ represents a more accurate estimation. Note that the accurate estimation of $\sigma_{Y}$ using all the PCE coefficients in Eq.~\eqref{eq:mom} indicates that the PCE represents an accurate spectral decomposition of the uncertainty in $Y$ with respect to $\boldsymbol{X}$. Thus, the PCE is useful for uncertainty quantification such as the variance-based sensitivity analysis, which aims to quantify the influence of each input on the output variance \cite{liu2018data}. 

In addition, we also compare the goodness of fit of two methods by estimating the KL divergence from the estimated output distribution to the true distribution \cite{boltz2009high}. Note that the relative error $\epsilon_{re}$ evaluates how well the two methods estimate the \textit{second-order moment} of $Y$ (which is important for sensitivity analyses in practice), whereas the KL divergence evaluates how well the methods approximate the \textit{overall distribution} of $Y$. These two metrics are also used to evaluate PCE models in \cite{torre2019data}.


\subsection{Ishigami function approximation}
We use the Ishigami function \cite{ishigami1990importance} in Eq.~\eqref{eq:ishigami} as our first simulation example to validate the proposed method for a model with \emph{independent} inputs.
\begin{equation}
\label{eq:ishigami}
    Y = \sin(X_{1}) + 7\sin^{2}(X_{2}) + 0.1X_{3}^4\sin(X_{1}), 
\end{equation}
where $X_{i} \sim \mathcal{U}(-\pi, \pi), i = 1, 2, 3$. This function is widely used as a test function to benchmark PCE methods due to its strong non-linearity and non-monotonicity \cite{torre2019data}.

In this example, we first show how to choose the threshold value for the proposed method. In addition, we compare the proposed method with the benchmark method in Section~\ref{subsec:sparsePCE} in terms of the estimation accuracy and computational efficiency.

\begin{figure}[h]
\begin{minipage}[t]{0.49\linewidth}  
\includegraphics[width=\linewidth]{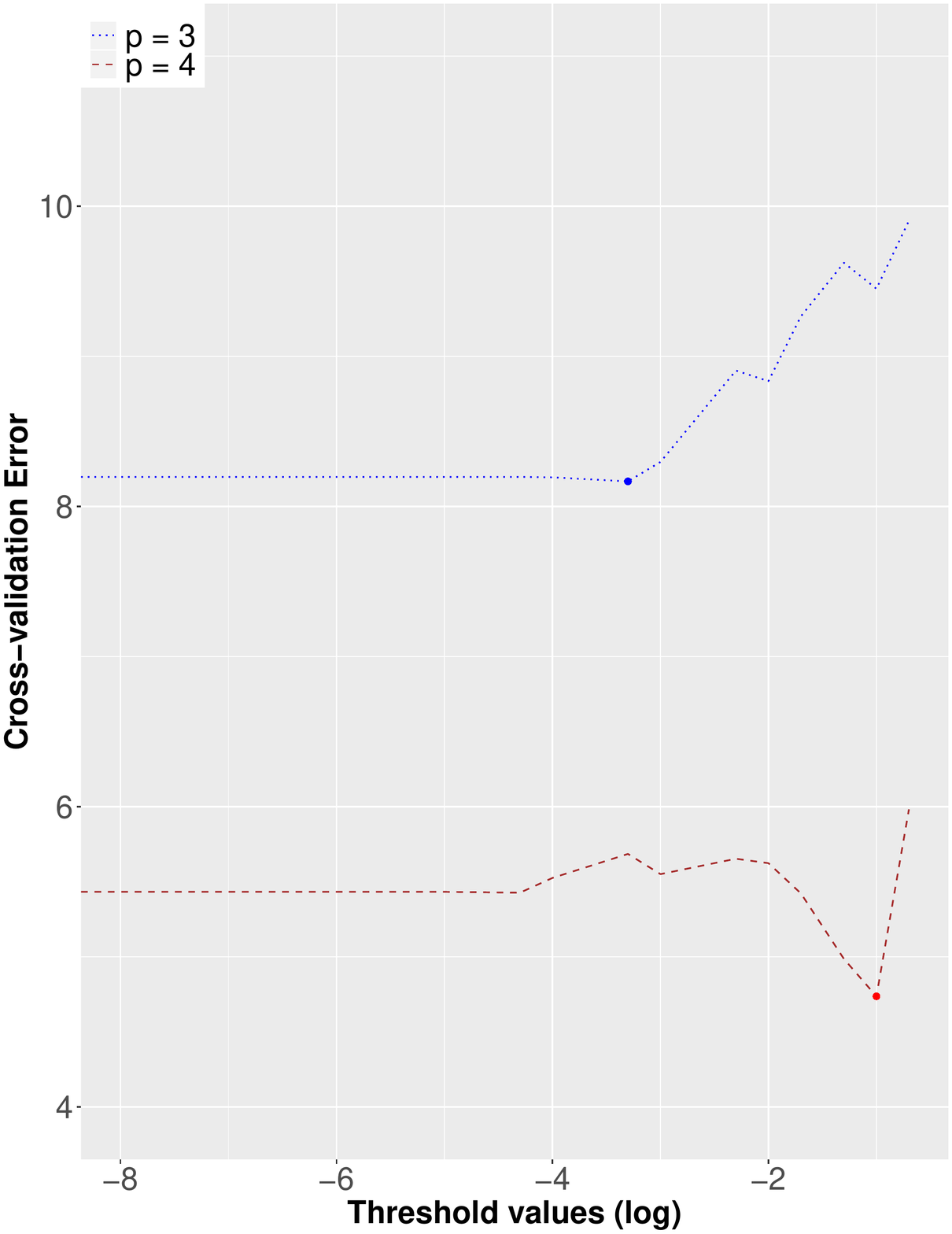} 
\end{minipage}
\hfill
\begin{minipage}[t]{0.49\linewidth}  
\includegraphics[width=\linewidth]{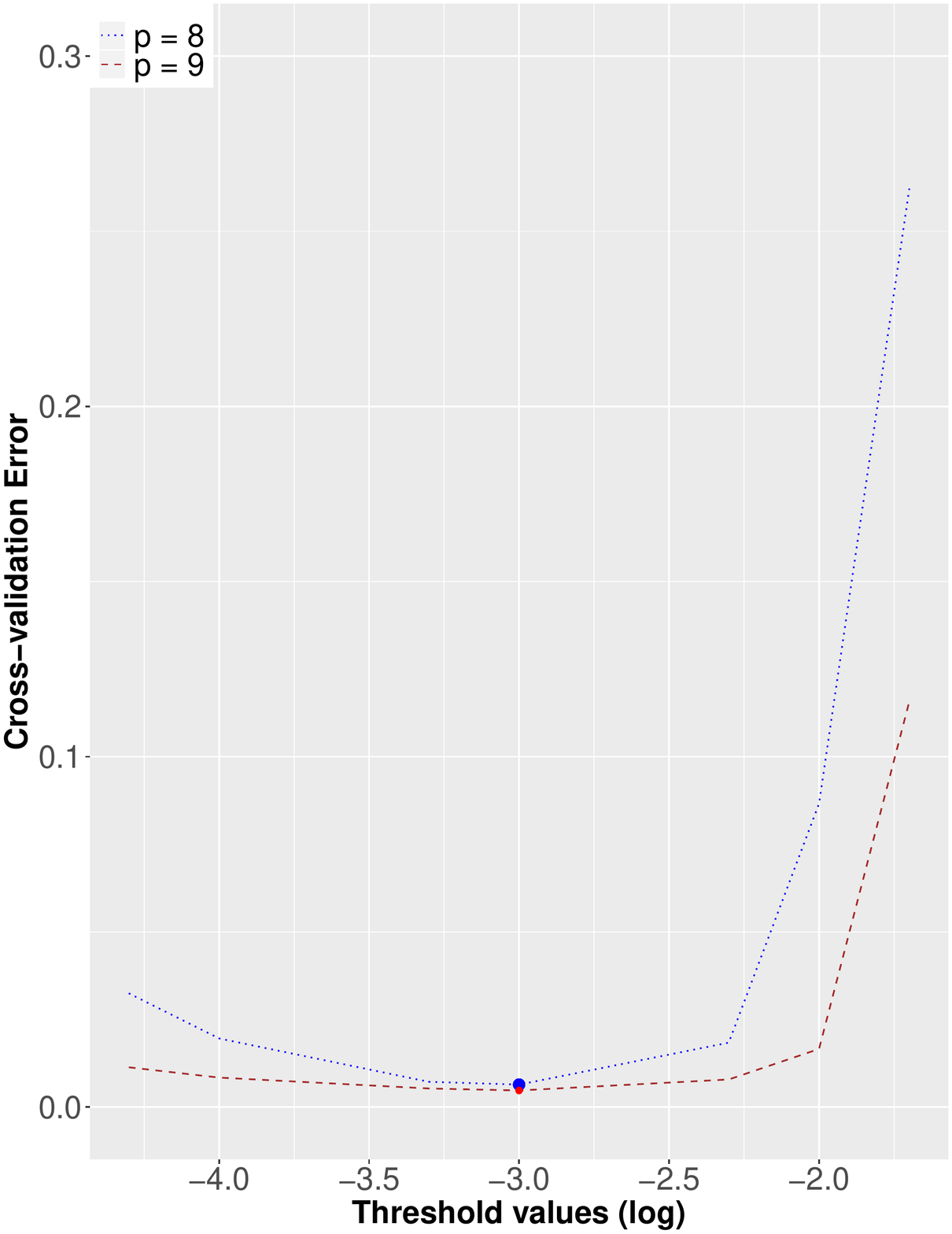}
\end{minipage}
\caption{The left subgraph shows the $5$-fold cross-validation errors of PCEs with $p=3$ and $p=4$ across different threshold values for the proposed method. The right subgraph shows the $5$-fold cross-validation errors of PCEs with $p=8$ or $p=9$ across different threshold values. The blue (resp. red) points represent the threshold values that correspond to the threshold values in Eq.~\eqref{eq:cross-vali} for PCEs with $p=3$ (resp. 4) or $p=8$ (resp. 9). }
\label{fig:ishigami_thres}
\end{figure}

The $5$-fold cross-validation is used to find the optimal threshold value for the proposed method considering different polynomial orders using $200$ random observations. The blue (resp. red) points of the left subfigure and right subfigure in Figure~\ref{fig:ishigami_thres} represent the threshold values that correspond to the minimal cross-validation errors as defined in Eq.~\eqref{eq:cross-vali} for the PCE with the polynomial order of 3 (resp. 4) and 8 (resp. 9), respectively. In addition, as it is presented in Figure~\ref{fig:ishigami_thres}, the proposed model with a higher polynomial order achieves a smaller cross-validation error than the model with a lower polynomial order. It reflects the fact that a more complex PCE model tends to better approximate a target function. 

\begin{figure}[h]
\begin{minipage}[t]{0.48\linewidth}  
\includegraphics[width=\linewidth]{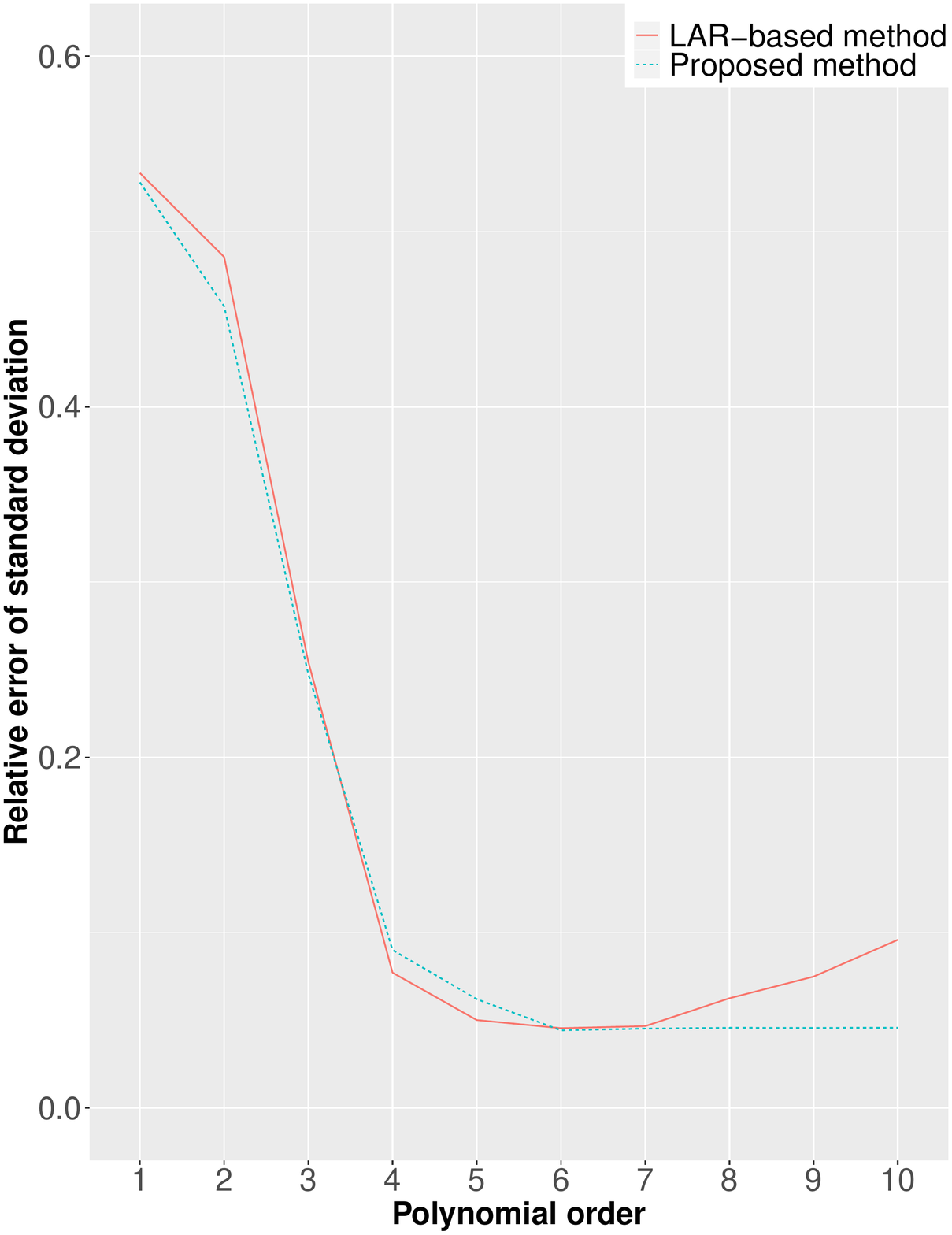} 
\end{minipage}
\hfill
\begin{minipage}[t]{0.48\linewidth}  
\includegraphics[width=\linewidth]{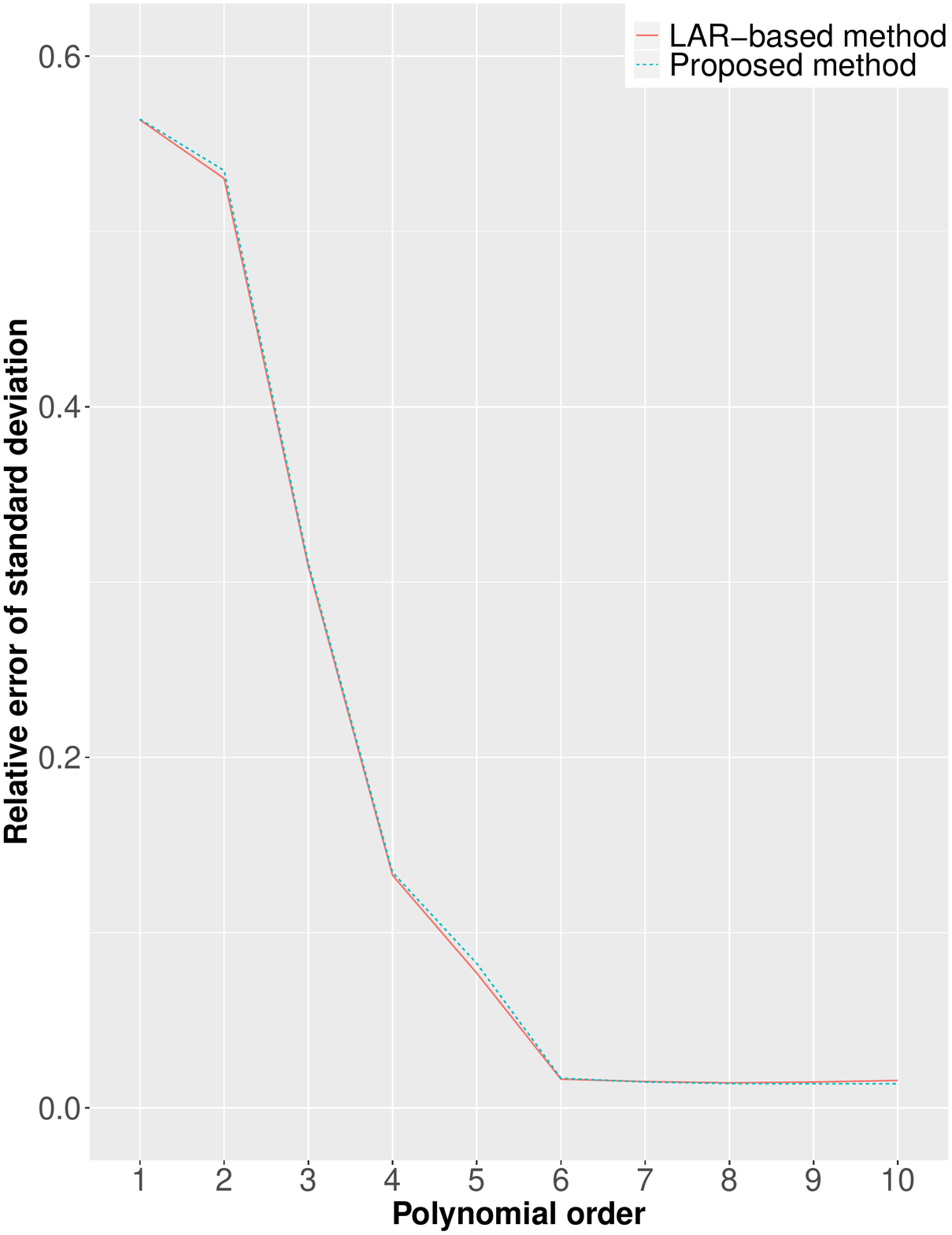}
\end{minipage}
\caption{The plots show the relative errors of estimating the standard deviation of $Y$ using $m=100$ (left) and $1000$ (right) random observations for both methods. 
The relative error is averaged across $50$ simulation runs for each polynomial order $p$. }
\label{fig:ishigami_error}
\end{figure}

We first compare how the polynomial order $p$ and the sample size $m$ interactively affect the performance for both the benchmark method and the proposed method. The left subfigure in Figure~\ref{fig:ishigami_error} shows that the proposed method achieves a better accuracy by increasing the polynomial order when the sample size is small. On the other hand, the performance of the benchmark method does not improve as the polynomial order increases when the polynomial order is greater than $6$. This is due to the numerical instability by the over-parametrization of using the GS-PCE based on an insufficient number of random observations. The right subfigure in Figure~\ref{fig:ishigami_error} shows the trend that increasing the polynomial order improves the accuracy for both methods given a large sample size. Therefore, we conclude that the proposed method achieves the similar or a better accuracy than the benchmark method given the same number of observations for models with independent inputs.

\begin{figure}[h]
\begin{minipage}[t]{0.48\linewidth}  
\includegraphics[width=\linewidth]{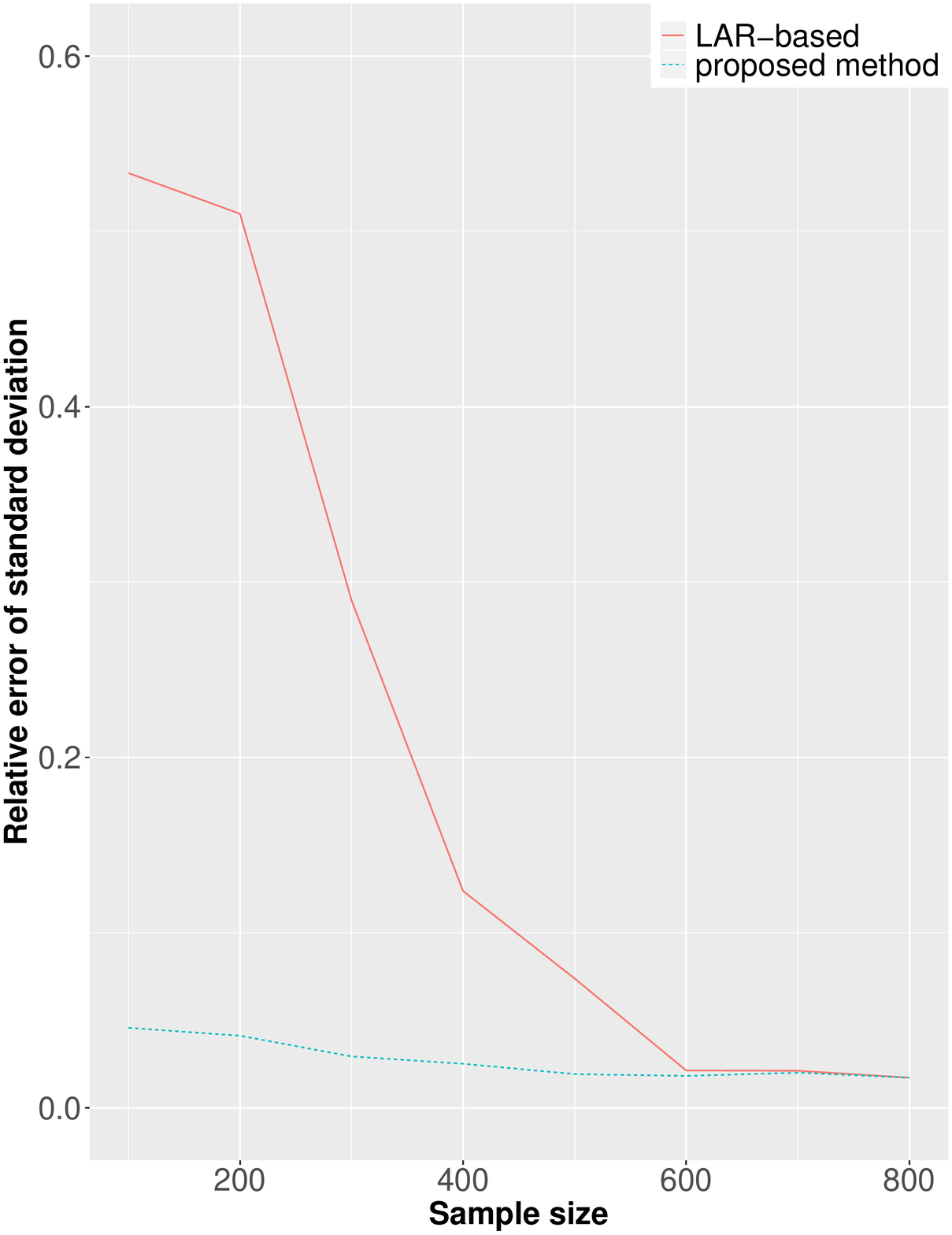} 
\caption{Relative errors of estimating the output standard deviation with $p=8$ v.s. the number of random observations. The relative errors are averaged across $50$ simulation runs for each sample size.}
\label{fig:ishigami_size}
\end{minipage}
\hfill
\begin{minipage}[t]{0.48\linewidth}  
\includegraphics[width=\linewidth]{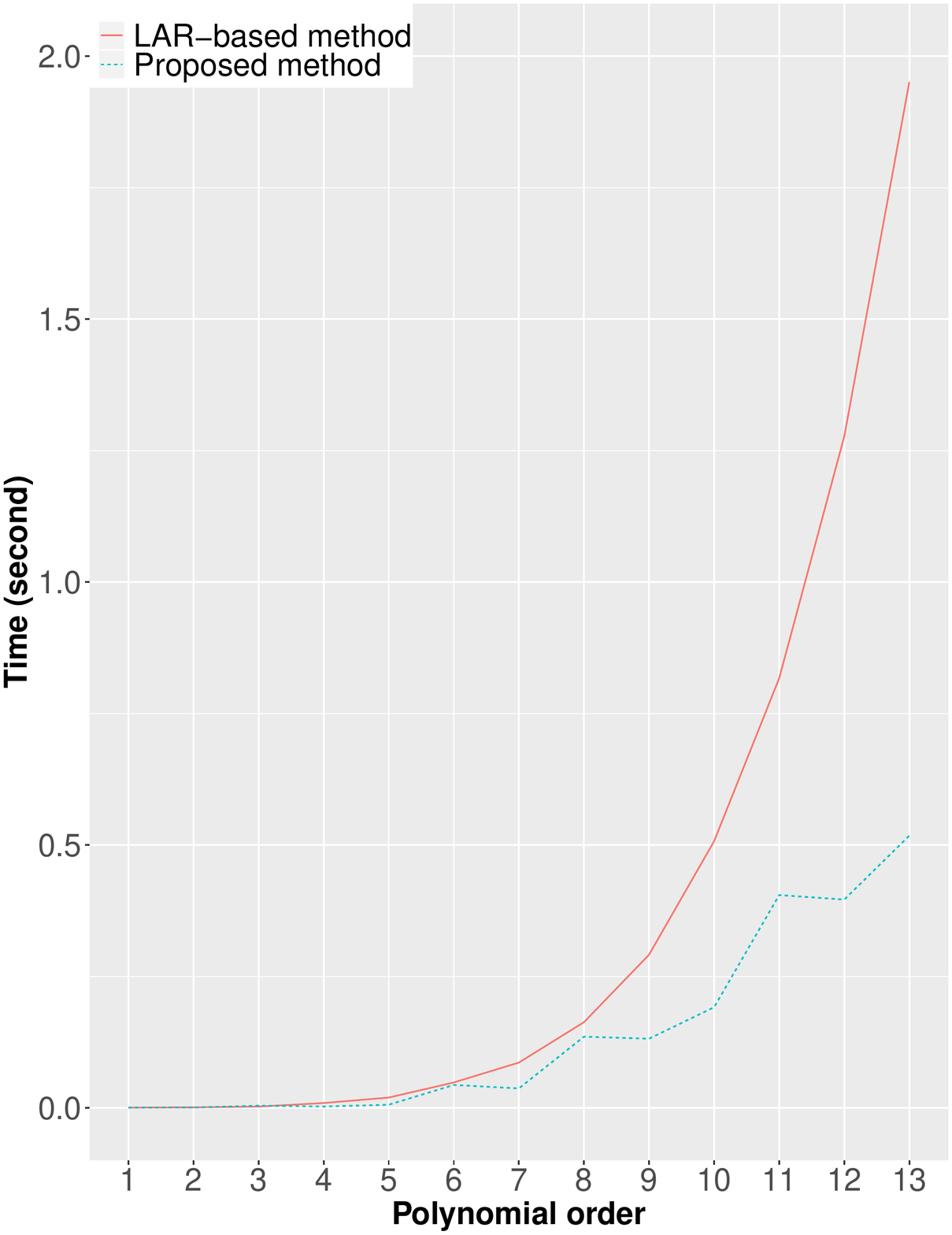}
\caption{Computational time (seconds) v.s. the polynomial order $p$ for both methods. For each polynomial order, the computational time is averaged across $50$ simulation runs, where each simulation run uses $1,000$ random observations. }
\label{fig:ishigami_time}
\end{minipage}
\end{figure}

Furthermore, we study how the sample size $m$ affects the estimation accuracy of the proposed method compared with the benchmark method based on a fixed polynomial order $p=8$. As shown in Figure~\ref{fig:ishigami_size}, the proposed method achieves a much better accuracy than the benchmark method when the sample size is small. When the sample size is large, both methods perform similarly as expected. Besides comparing the estimation accuracy of the standard deviation, we also compare the model performance by estimating the KL divergence using two different methods on different sample sizes. When the sample size is $1000$, the mean and standard error of the KL divergence for the proposed method and the benchmark method are $-0.0031$ $(\pm 0.0003)$ and $0.0057$ $(\pm 0.0004)$ based on $50$ simulation runs, respectively. When the sample size is $100$, the mean and standard error of the KL divergence using the proposed method is $0.005$ $(\pm 0.001)$. It is much smaller than the KL divergence of using the benchmark method $0.055$ $(\pm 0.003)$. This also validates that the proposed method has a better computational accuracy than the benchmark method.

In addition, we compare the computational efficiency of the proposed method with the benchmark method in terms of the computational time. 
The computational times are recorded using a 1.4 GHz Intel Core i5 machine with a 16 GB 1600 MHz DDR3 RAM. The average computation times for both methods are calculated based on $50$ simulation runs for the polynomial order of $p=1$ through $p=13$ using $1,000$ random observations in each simulation run. As shown in Figure~\ref{fig:ishigami_time}, the computational time for the benchmark method increases exponentially as the polynomial order increases. It can be explained by Eq.~\eqref{eq:exp2} since the number of constructed orthonormal polynomials increases exponentially as the polynomial order $p$ increases. However, the computation time for the proposed method grows much more slowly. This can be intuitively explained by Step 2 in Algorithm \ref{algorithm_sf} since the number of polynomials is reduced in each iteration.


\subsection{Numerical example with dependent inputs}
We use a numerical example in \cite{liu2018data} as our second example to validate the proposed method for models with dependent inputs. This example involves multiple types of probability distributions of inputs as follows:
\begin{equation}
\begin{aligned}
\label{eq:sim2formula}
&\begin{pmatrix}
X_{1}\\
X_{2}\\
X_{3}\\
X_{4}
\end{pmatrix} & \sim & \mathcal{N} \left[\left(\begin{array}{c}
0\\
0\\
0 \\
0
\end{array}\right),\left(\begin{array}{cccc}
1 & 0 & 0 & 0\\
0 & 1 & 0 & 0\\
0 & 0 & 1 & 0.3\\
0 & 0 & 0.3 & 1\\
\end{array}\right)\right], \\
& X &\sim &\mathcal{U}(0,\,1), \\ 
& X_{5} &= &\theta_{1}X + \mathcal{U}(0,\,1), \\
& X_{6} &= &\theta_{2}X + \theta_{3}X^{2} + \mathcal{U}(0,\,1), \\
&Y &=& X_{1}X_{2} + X_{3}X_{4} + X_{5}X_{6}.
\end{aligned}
\end{equation}
Here, we set $(\theta_1,\theta_2,\theta_3) = (0.4, 0.6, 1)$ as in \cite{liu2018data}. We also compare the estimation accuracy of the standard deviation of $Y$ using the two methods. Unlike the first example, where it requires a PCE model with a large $p$ to model the response function, this example uses a PCE model with $p=2$ for both methods. We measure the performance of each method using $m=20$ and $m=100$ and report the results based on $50$ replications. 
As it is shown in Table \ref{tab:num2}, the proposed method achieves the same accuracy as the benchmark method when $m=100$. However, the proposed method provides a much better accuracy than the benchmark method when $m=20$. In addition, as it is shown in Table \ref{tab:num2} the estimated KL divergence of using the benchmark method is infinity when $m=20$. It shows that the benchmark method cannot model the input-output relationship. However, the small KL divergence of using the proposed method indicates the robustness of the proposed method. When $m=100$, the proposed method still shows a better estimation of the output distribution than the benchmark method.   


\begin{table}[H]
\caption{Estimations of the output standard deviation and KL divergence using the benchmark method and the proposed method across $50$ simulation runs. Each simulation run uses $m=20$ or $m=100$. The relative errors are calculated based on the theoretical value $\sigma_{Y} = 1.655$ provided in \cite{liu2018data}. The estimation accuracy of the proposed method is better than the benchmark method when $m=20$. The KL divergence of the proposed method is smaller than the benchmark method for both sample sizes. } 
\label{tab:num2}
\centering
\begin{adjustbox}{width=0.98\textwidth}
\small
\begin{tabular}{|c|c|c|c|c|}
\hline
Sample size          & Method           & Estimation      & Relative error& KL divergence \\ \hline
\multirow{2}{*}{20}  & Benchmark method & $0.914 \pm 0.112$ & $44.77\%$ &   $\infty$   \\  \cline{2-5} 
                     & Proposed method  & $1.618 \pm 0.051$ & $2.23\%$  &  $0.087 \pm 0.013$ \\ \hline
\multirow{2}{*}{100} & Benchmark method & $1.643 \pm 0.027$ & $0.73\%$  &   $0.016 \pm 0.002$  \\ \cline{2-5} 
                     & Proposed method  & $1.643 \pm 0.027$ & $0.73\%$ &   $0.007\pm0.001$  \\ \hline
\end{tabular}
\end{adjustbox}
\end{table}

\subsection{23-bar horizontal truss}
\begin{figure}[h]
\includegraphics[width=\linewidth]{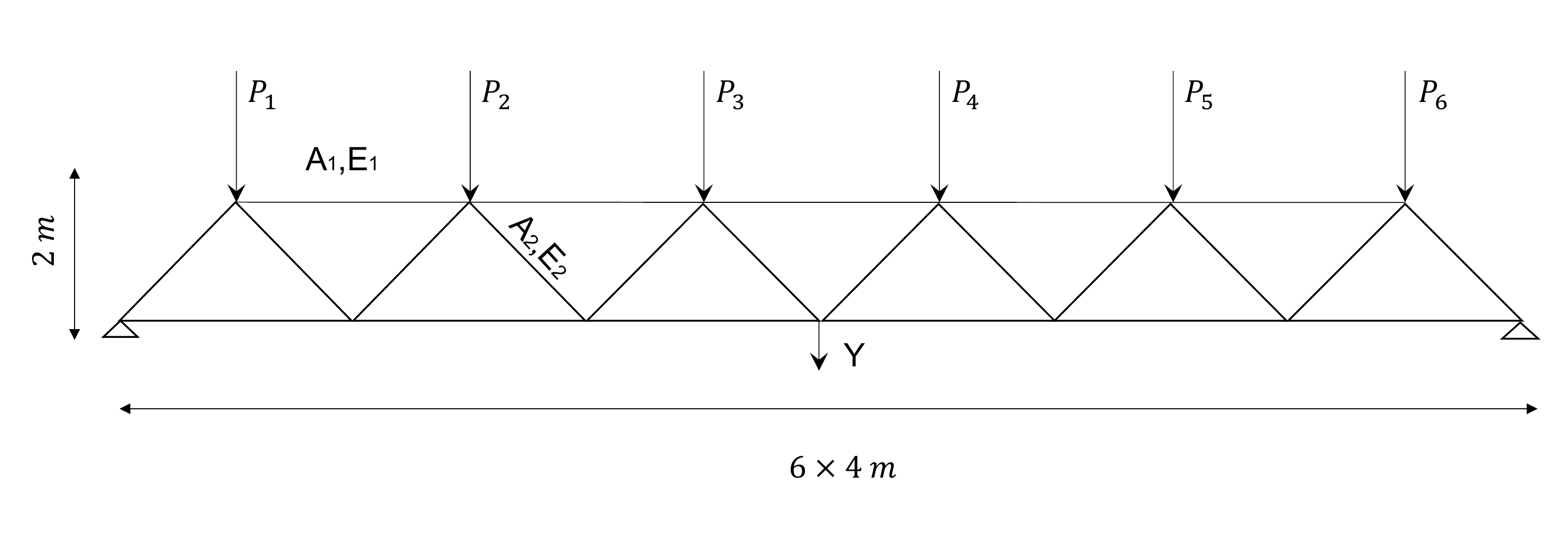} 
\caption{Schema of the horizontal truss model modified from \cite{liu2018data}. Young modulus $E_{i}, i=1,2$, cross-sectional area $A_{i}, i=1,2$ for horizontal and diagonal bars, and the random loads $P_{i}, i=1,2,\cdots,6$ are the inputs which affect the downward vertical displacement at the mid span of the structure, $Y$.}
\label{fig:truss}
\end{figure}

We consider the 23-bar horizontal truss example in \cite{torre2019data} as our third example. The downward vertical displacement at the mid span of the structure, $Y$, is considered as the output of interest. As depicted in Figure~\ref{fig:truss}, the uncertainty of $Y$ is affected by Young modulus $E_{i}, i=1,2$, cross-sectional area $A_{i}, i=1,2$ for horizontal and diagonal bars, and the random loads $P_{i}, i=1,2,\cdots,6$. All inputs in this example have the same distributions as in \cite{torre2019data}. $E_{i}, i=1,2$ and $A_{j}, j=1,2$ are assumed to be mutually independent inputs and following the lognormal distribution with mean $\mu$ and standard deviation $\sigma$ as follows:
\begin{equation}
    \label{eq:truss}
    \begin{aligned}
    E_{1}, E_{2} \sim \mathcal{LN}(2.1 \times 10^{11}, 2.1\times 10^{10}) \; \textrm{[Pa]},\\
    A_{1} \sim \mathcal{LN}(2.0 \times 10^{-3}, 2.0 \times 10^{-4}) \; [\textrm{m}^2],\\
    A_{2} \sim \mathcal{LN}(1.0 \times 10^{-3}, 1.0 \times 10^{-4}) \; [\textrm{m}^2].
    \end{aligned}
\end{equation}

Unlike $E_{i}, i =1,2$ and $A_{i}, i=1,2$ which are mutually independent, $P_{i}, i=1,2,\cdots,6$ are mutually dependent on each other. In addition, $P_{i}$ marginally follows a Gumbel distribution with mean $\mu = 5\times 10^{4} \;\textrm{[N]}$ and standard deviation $\sigma = 7.5\times 10^{3} \;\textrm{[N]}$ with the marginal cumulative distribution function as follows:
\begin{equation}
    F_{i}(x;\alpha, \beta) = e^{-e^{-(x-\alpha)}/\beta}, i =1,2,\ldots,6,
\end{equation}
where $\beta = \sqrt{6}\sigma/\pi, \alpha = \mu - \gamma \beta$, and $\gamma \approx 0.5772$ is the Euler-Mascheroni constant. The dependency among $P_{i}, i=1,2,\ldots,6$ is encoded using the C-vine copula with the density as follows:
\begin{equation}
\label{eq:corre}
c_{\boldsymbol{X}}^{(\mathcal{G})}(u_{1},\ldots,u_{6}) = \prod_{j=2}^{6}c_{1j;\theta=1.1}^{(\mathcal{G}\mathcal{H})}(u_{1},u_{j}),    
\end{equation}
where $c_{1j;\theta=1.1}^{(\mathcal{G}\mathcal{H})}$ is the density of the pair-copula between $P_{1}$ and $P_{j}$, $j=2,\ldots,d$. $\mathcal{G}\mathcal{H}$ represents the Gumbel-Hougaard family whose bivariate copula can be represented as follows:
\begin{equation}
\begin{aligned} \nonumber
    C^{(\mathcal{G}\mathcal{H})}_\theta(u,v) =\exp\left(-\left( \left(-\log u\right)^\theta + \left(- \log v\right)^\theta \right)^{1/\theta}\right), \quad \theta \in \left[1, \infty\right),
 \end{aligned}
\end{equation}
where ${\theta}$ decides the correlations among the loads. Based on Eq.~\eqref{eq:corre}, we can see that $P_{1}$ is equally correlated with all the other loads. $Y$ is simulated based on a regression of the standardized inputs with coefficients provided in \cite{lee2006response} as follows:
\begin{equation}
\begin{aligned}
Y &= 2.8070 + 1.2598E^{\prime}_{1} + 0.2147E^{\prime}_{2} + 1.2559A^{\prime}_{1} + 0.2133A^{\prime}_{2} - 0.1510P^{\prime}_{1} - 0.4238P^{\prime}_{2} - \\ &0.6100P^{\prime}_{3} - 0.6100P^{\prime}_{4} -0.4238P^{\prime}_{5} - 0.1510P^{\prime}_{6} - 0.1978E^{\prime2}_{1} - 0.0362E^{\prime2}_{2} - 0.2016A^{\prime2}_{1} - \\ & 0.0346A^{\prime2}_{2}+0.0023P^{\prime2}_{1}+0.0008P^{\prime2}_{2}+0.0036P^{\prime2}_{3}+ 0.0036P^{\prime2}_{4}+0.0008P^{\prime2}_{5}+0.0023P^{\prime2}_{6}- \\
& 0.0042E^{\prime}_{1}E^{\prime}_{2} -0.3022E^{\prime}_{1}A^{\prime}_{1}-0.0110E^{\prime}_{1}A^{\prime}_{2}+0.0381E^{\prime}_{1}P^{\prime}_{1}+0.0871E^{\prime}_{1}P^{\prime}_{2}+0.1232E^{\prime}_{1}P^{\prime}_{3}+ \\
& 0.1232E^{\prime}_{1}P^{\prime}_{4}+0.0871E^{\prime}_{1}P^{\prime}_{5}+0.0346E^{\prime}_{1}P^{\prime}_{6}+0.0041E^{\prime}_{2}A^{\prime}_{1}+0.0110A^{\prime}_{1}A^{\prime}_{2}+0.0261A^{\prime}_{1}P^{\prime}_{1}+ \\
& 0.0831A^{\prime}_{1}P^{\prime}_{2}+0.1172A^{\prime}_{1}P^{\prime}_{3} + 0.1172A^{\prime}_{1}P^{\prime}_{4} + 0.0832A^{\prime}_{1}P^{\prime}_{5} + 0.0296A^{\prime}_{1}P^{\prime}_{6},
    \end{aligned}
\end{equation}
where $E^{\prime}_{i}, i =1,2$, $A^{\prime}_{i}, i=1,2$, and $P^{\prime}_{i}, i=1,2,3,4,5,6$ are the standardized inputs. For example, $E^{\prime}_{1} = \frac{E - \mu_{E_{1}}}{\sigma_{E_{1}}}$, where $\mu_{E_{1}}$ is the mean of $E_{1}$ and $\sigma_{E_{1}}$ is the standard deviation of $E_{1}$.

We evaluate the benchmark method and the proposed method by averaging their performances over $50$ simulation runs, where each simulation run uses $m=20$ or $m=100$. As shown in Table \ref{tab:truss}, the proposed method attains essentially the same performance as the benchmark method when $m=100$. However, the proposed method achieves a much better accuracy when $m=20$ in terms of both the standard deviation and KL divergence.

\begin{table}[H]
\caption{Estimations of the output standard deviation and KL divergence using the benchmark method and the proposed method based on $50$ simulation runs. Each simulation run uses $m=20$ or $m=100$. The relative errors are calculated based on $\sigma_{y} = 2.169$, where it is estimated based on a Monte Carlo estimator with $100$ simulation runs, each of which uses $10^{5}$ random observations. The proposed method shows a better estimation accuracy than the benchmark method when $m=20$. } 
\label{tab:truss}\centering
\begin{adjustbox}{width=0.98\textwidth}
\small
\begin{tabular}{|c|c|c|c|c|}
\hline
Sample size          & Method           & Estimation      & Relative error & KL divergence\\ \hline
\multirow{2}{*}{20}  & Benchmark method & $ 2901.137\pm  2899.45 $ & $>100\%$ &  $\infty$   \\  \cline{2-5} 
                     & Proposed method  & $ 2.039 \pm  0.059 $ & $5.99\%$  & $0.113\pm0.008$    \\ \hline
\multirow{2}{*}{100} & Benchmark method & $ 2.156\pm  0.029 $ & $0.61\%$ &   $0.018\pm0.002$   \\ \cline{2-5} 
                     & Proposed method  & $ 2.156\pm  0.029 $ & $0.61\%$ & $0.018\pm0.003$    \\ \hline
\end{tabular}
\end{adjustbox}
\end{table}

\subsection{HIV model}
The HIV model used in \cite{zhu2018analytic} is considered as our fourth example to validate the proposed method. The output of interest is the basic reproduction number $(R_{0})$, which is arguably regarded as the most important quantity that measures the effectiveness of an infectious disease spreading through a population \cite{fraser2009pandemic,holme2015basic}. $R_{0}$ is modeled using a deterministic equation as follows:
\begin{equation}
    \label{eq:HIV1}
    R_{0} = \frac{\beta_{0}(1-\gamma)\theta_{d}^{2} + \beta_{1}n_{1}Q_{0}(n_{d}-\kappa) + \beta_{2}n_{2}\alpha Q_{0} + (1-\gamma)(\kappa + \alpha) \beta_{0}\theta_{d}}{\theta_{d}(\theta_{d}+\kappa)(\theta_{d}+\alpha)},
\end{equation}
where the inputs follow uniform distributions with the parameters listed in Table \ref{tab:HIVpara}. In addition, there exist correlations between $\beta_{1}$ and $n_{1}$ as well as between $\beta_{2}$ and $n_{2}$, where the Pearson correlation coefficients are $\rho_{\beta_{1}, n_{1}} = 0.3$ and $\rho_{\beta_{2}, n_{2}} = 0.5$, respectively.  

\begin{table}[H]
\caption{Input descriptions and distributions of the HIV model.} 
\label{tab:HIVpara}\centering
\begin{adjustbox}{width=0.98\textwidth}
\small
\begin{tabular}{|c|c|c|}
\hline
Input        & Descriptions & Distribution       \\
\hline
$Q_{0}$      & Recruitment rate & $U(0.0261,0.0319)$ \\
$\beta_{0}$  & Birth rate of infective& $U(0.027, 0.033)$  \\
$\gamma$     & Fraction of susceptible newborn from infective class& $U(0.36, 0.44)$    \\
$\beta_{1}$  &Contact rate of susceptible with asymptomatic infective& $U(0.18, 0.22)$    \\
$\beta_{2}$  &Contact rate of susceptible with symptomatic infective& $U(0.072, 0.088)$  \\
$n_{1}$      &Number of sexual partners of susceptible with asymptomatic infective& $U(1.8, 2.2)$      \\
$n_{2}$      &Number of sexual partners of susceptible with symptomatic infective& $U(1.8, 2.2)$      \\
$\theta_{d}$ &Natural death rate& $U(0.018, 0.022)$  \\
$\alpha$     &Removal rate of symptomatic class& $U(0.54,0.66)$     \\
$\kappa$     &Rate of development to AIDs& $U(0.09,0.11)$    \\
\hline
\end{tabular}
\end{adjustbox}
\end{table}

In this example, we set $p=4$ for both the benchmark method and the proposed method. The benchmark method requires at least $2,000$ random observations for the $10$ random inputs to keep the orthogonality of the constructed orthonormal polynomials using the modified Gram-Schmidt algorithm. The lack of orthogonality of the constructed orthonormal polynomials causes inaccurate estimation, as shown in Table \ref{tab:hiv2} for $m$=200. However, the proposed method still achieves accurate estimation for the same sample size. This also reflects the fact that the proposed method reduces the number of random observations to construct a sparse PCE for models with dependent inputs. In addition, the KL divergence shown in Table \ref{tab:hiv2} suggests that the proposed method has a much better performance on modeling the relationship between the input and the output.


\begin{table}[H]
\caption{Estimations of the output standard deviation and KL divergence using the benchmark method and the proposed method across $50$ simulation runs. Each simulation run uses a $m=200$. The relative errors are calculated based on the theoretical value $\sigma_{Y} = 0.252$ provided in \cite{zhu2018analytic}. The proposed method provides more accurate estimation than the benchmark method. } 
\label{tab:hiv2}\centering
\begin{adjustbox}{width=0.98\textwidth}
\small
\begin{tabular}{|c|c|c|c|c|}
\hline
 Method   &Sample size                   & Estimation      & Relative error &KL divergence\\ \hline
 Benchmark method &\multirow{2}{*}{200}   & $1034.934\pm 585.089$ & $> 100\%$ &  $\infty$   \\  \cline{1-1} \cline{3-5} 
Proposed method  &                     & $0.258  \pm0.002$ & $2.33\%$ &  $0.013\pm 0.002$    \\ \hline
\end{tabular}
\end{adjustbox}
\end{table}

\section{Conclusion and future work}
\label{sec:conclusion}
In this paper, we propose a data-driven sparse PCE for models with dependent inputs without requiring any distributional information about the inputs or a large number of random observations. Four numerical examples are used to validate the proposed method. It is shown that the proposed method accurately estimates the standard deviation and distribution of the output using a small sample size and improves upon the computational efficiency of the benchmark method for constructing a sparse PCE.

A recent work \cite{liu2018data} provides interpretable sensitivity indices for models with dependent inputs without assuming the distributions or dependence structures of the inputs. 
This suggests a future research direction on estimating the sensitivity indices proposed in \cite{liu2018data} using the proposed sparse PCE. In addition,
the proposed method has limitations in searching all possible thresholds. Coming up with a way to efficiently find the threshold for the proposed method (analogous to the LARS algorithm for fitting all possible LASSO models on data) is a future research direction.


\bibliography{mybibfile}

\end{document}